\documentstyle[epac]{article}
\begin{document}
\bibliographystyle{unsrt}
\title{ Comparisons Between Modeling and Measured Performance of the BNL Linac$^*$}
\author{D.\ Raparia, J. G. \ Alessi, A. \ Kponou\\
        Brookhaven National  Laboratory \\
        Upton, NY, 11973, USA \\ }
\thanks{Work performed under the auspices of the U. S. Department of Energy.}
\maketitle

\begin{abstract}
Quite good agreement has been achieved between computer modeling
and actual performance of the Brookhaven 200 MeV Linac. We will
present comparisons between calculated and measured performance
for beam transport through the RFQ, the 6 m transport from RFQ to 
the linac, and matching and transport through the linac.
\end{abstract}

\section{Introduction}			

	The Brookhaven 200 MeV linac serves as the injector 
	for the AGS Booster and as well  delivers beam to 
	the Brookhaven Isotope Resource Center.
	It consists of 
	a 35 keV magnetron surface plasma  source,
	a low energy beam transport (LEBT) \cite {lebt},
	201 MHz radio frequency quadrupole (RFQ) \cite {RFQ},
	medium energy beam transport (MEBT) 
	and
	200 MeV Linac \cite {200mevlinac}.
	In this last year we have gone through  a linac upgrade to 
	get 2.5 more average current
	(146 $\mu$ A) \cite {upgrade}. This was achieved 
	by increasing repetition rate 5 to 7.5 Hz and increasing 
	peak current from 25  to 39 mA.
	In this paper we compare computer modeling 
	with actual performance.

\section{LEBT and RFQ}

	LEBT had two  pulsed solenoids, two sets of x and y steerer, beam
   	chopper, emittance probe, and two current toroids. The chopper
	was removed  from
	the line, making the line   shorter by 70 cm.
	Computer modeling of this line showed that we should move the 1st
	solenoid as close to the ion source  as possible
	to reduce the beam size in the 1st solenoid and the second
	solenoid as close to the RFQ as possible to increase the convergence
	angle required by the RFQ  acceptance. 
	Figure 1 shows the
	ion trajectories through this line and phase space at the exit of
	the ion source, middle of the line and entrance of RFQ. 
	The RFQ acceptance is shown as a solid line ellipse.
        This calculation  assumed that the  beam space charge is 
	neutralized.
\begin{figure}
	\vspace{2.0in}
	\includegraphics{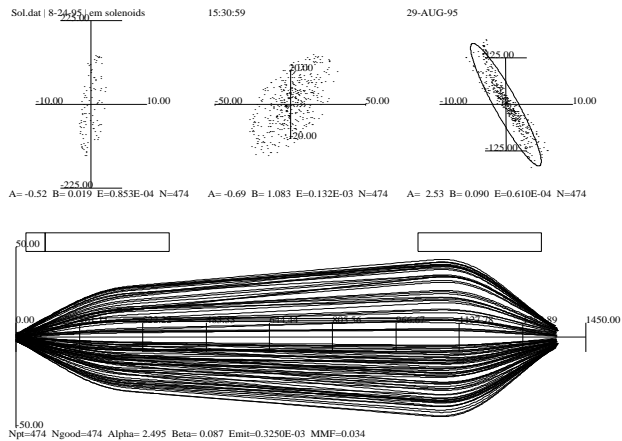}
        \hspace{.15in}
	\caption{ Ion trajectories through the LEBT and phase space
	at ion-source, middle of the line, and at the entrance  of the RFQ.}
\end{figure}

	Shortening of the line resulted in lower measured emittance.  
        Due to lower emittance
	and better matching  transmission  through the  RFQ was improved
	by about 10 percent.
	Figure 2 shows the transmission as a function of input beam current.	
\begin{figure}
	\vspace*{1.0in}
	\includegraphics{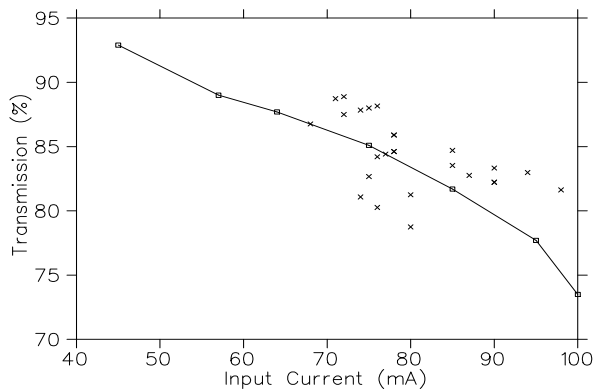}
	\caption{ RFQ transmission as a function of input current.
	The solid line indicates the simulation and crosses  the measured
	values.}
\end{figure}

\section{MEBT and Linac }
	This is where we have recovered most of the beam losses.
	This line is 6 meter long and is shown in Figure 3. It consists of
        four triplets, three bunchers, one slow chopper, one fast 
	chopper, two emittance measurement units, three current
	transformers, two sets of x, y steerers, and a dipole to accomodate
	polarized beam coming at 60 deg angle. The ideal match between RFQ and
	DTL could have been  obtained with a 5 \( \beta \lambda \) long FODO 
	lattice with quadrupole spacing  about  \(\beta \lambda  \)
 	and at least two bunchers. But the requirement of beam chopping and
	polarized beam dictated a triplet solution \cite {rfqdtlmatch}.
\begin{figure}
	\vspace*{2.5in}
	\includegraphics{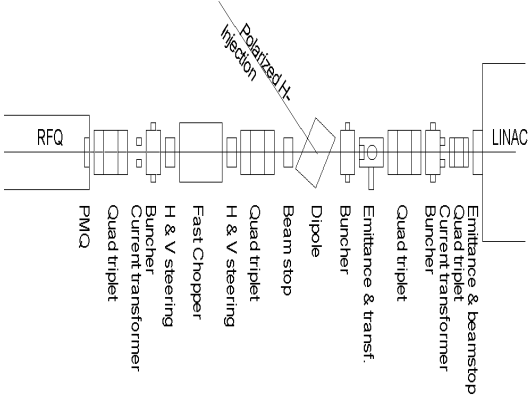}
	\caption{ MEBT, 750 keV Transport Line.}
\end{figure} 

	The first quadrupole after the RFQ was too far; by the time 
	the beam reached the quadrupole it had gone
	through a  waist in the x  plane, hence was diverging in both planes.
        No matter which  polarity quad one puts, beam size in  
	the other direction 
	is very big. 
	Also the longitudinal beam size is too big before it reaches
	the first buncher. 	
	To improve the capture and transmission of the beam in MEBT,
	the RFQ end flange at the high energy end was modified to accommodate 
	a permanent magnet quadrupole (PMQ).
	The PMQ was similar to one as used in the SSC DTL \cite {sscdtl}.
	We have also rearranged the gate valve and current transformers
	at the beginning of the line, and also measured and aligned all the
	quadrupoles very carefully.  Measurement as well as
	simulation showed  that as little as a 1.5 degree quadrupole
	rotation can increase the emittance by 50 \%. The last quadruplet
	was changed to a triplet to reduce coupling. Figure 4, shows
	the measured and calculated phase spaces after the second buncher.
	Table 1 shows  calculated (TRACE3D) and measured Twiss parameters for
	Figure 4.
\begin{figure}
	\vspace*{2.5in}
	\includegraphics{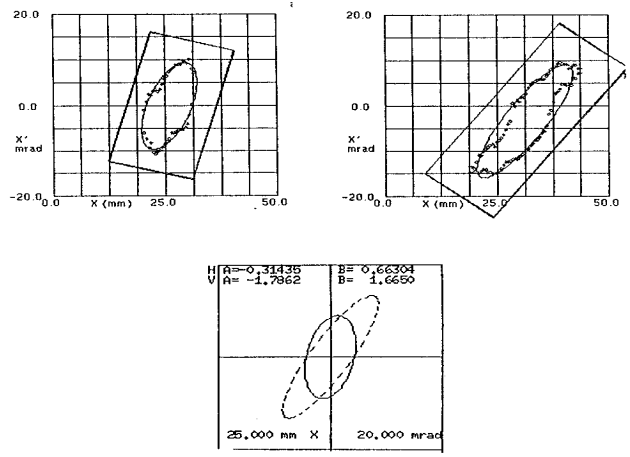}
	\caption{Measured (upper two) and calculated (lower) phase space 
	 after the second buncher. }
\end{figure} 
\begin{table}
 \begin{center}
	  \begin{tabular}{||l|l|l|l|l|l|l||}
	\hline
        \hline
		Planes 					&
                \multicolumn{3}{c|}{X}			&
                \multicolumn{3}{c||}{Y}			\\
		\cline{2-4} \cline{5-7}                         &
		\( \alpha_x \)					&
		\( \beta_x \)      				&
		\( \epsilon_x \)				&
		\( \alpha_y \)    				&
		\( \beta_y \)					&
		\( \epsilon_y \)        			\\
	\hline	Meas.						&
		-0.60						&
		 0.79						&
		 9.71						&
		-1.59						&
		 1.61						&
		 13.82						\\
	\hline
		Cal.						&
		 -0.31						&
		  0.66						&
		  9.80						&
		  -1.79					&
		  1.67						&
		  14.00						\\
	\hline  
	\hline
  \end{tabular}
 \end{center}
	\caption{Calculated and measured Twiss parameters after 
	the second buncher.
	\( \beta \) is in mm/mrad and \( \epsilon \) (unnor.,RMS) in \( \pi \) mm mrad. }
\end{table}

	Table 2, compares measurements and PARMILA results at various locations.
	We believe that lower values for currents after the second buncher and
	and at the entrance of the Linac are caused by the grids in
	the buncher drift tubes (four in  each buncher). These
	grids are placed in the bunchers to reduce RF defocusing effects. 
	Emittance measurements at 200 MeV are done using profiles at five
	places. Agreement between calculations and measured Twiss
	parameters at 200 MeV 
	is poor
	because there are 295 quadrupoles, and calibration and misalignment
	errors are not known to a good accuracy.
\begin{table}
 \begin{center}
	  \begin{tabular}{||l|l|l|l|l||}
	\hline
        \hline
		Location 					&
                \multicolumn{2}{c|}{94-95}			&
                \multicolumn{2}{c||}{95-96}			\\
		\cline{2-3} \cline{4-5}                         &
		Sim.					&
		Meas.      					&
		Sim.      					&
		Meas.        			   		\\
	\hline
                \multicolumn{5}{||c||}{Current mA}			\\
	\hline	 RFQ						&
		50.0						&
		  50.0						&
		  62.9						&
		  62.9						\\
	\hline
		 Buncher						&
		50.0						&
		  41.0						&
		  62.9						&
		  57.8						\\
	\hline
		MEBT					 	&
		42.4						&
		  37.3						&
		  62.9						&
		  53.2						\\
	\hline
                Tank 1      				          &
		27.4						&
		  28.4						&
		  37.1						&
                  37.7						\\
	\hline  
		Tank 9						&
		26.2						&
		  26.7						&
		  36.4						&
		  35.9						\\
	\hline
                \multicolumn{5}{||c||}{Emittance, (nor,RMS) $\pi$ mm mrad}\\
	\hline
		RFQ		    				&
		0.400						&
		   0.400						&
		    0.375						&
		     0.375						\\
	\hline
		Buncher						&
		0.44						&
		   0.56						&
		     0.47						&
		      0.57						\\
	\hline
		200 MeV						&
		1.32						&
		   2.8						&
		     1.85						&
		      1.92						\\
	\hline
	\hline
  \end{tabular}
 \end{center}
	\caption{Comparison between simulations and measured
	beam parameters.}
\end{table}

\section{Algebraic Reconstruction Technique (ART)}

 A radiograph of the beam at the BLIP target taken last year showed a
tilted ellipse in the x-y plane.
  Sources of this coupling can be quad rotation
or vertical offset in the dipole. This triggered the need for an x-y density
profile. We found that algebraic reconstruction
technique (ART) could help us. ART was introduced by Gordan, Bender and Herman
\cite {art} for solving the problem of three dimensional reconstruction
from projections. The ART algorithms have a simple intuitive basis.
Each projected density is thrown back across the reconstruction space in which
the densities are iteratively modified to bring each reconstructed projection 
into agreement with the measured projection.
The reconstruction space is an n x n array of small pixels, $\rho$ is grayness
or density number which is uniform within the pixel but different from
 other pixels.
Assume \bf{P} \rm is a matrix of m x n$^{2}$ and the m component column vector
\bf{R}. \rm Let $p_{i,j}$ denote the (i,j)th element of \bf{P} \rm, and $R_i$ demote
the ith element of reconstructed projection vector \bf{R}\rm.
 For $1 \leq i \leq m$, N$_i$ is number of pixels under projection R$_i$,
defined as
 $ N_{i} = \sum_j p_{i,j}^{2}$.
 The density number $ \rho_j^q $ denotes the value of $ \rho_j $ after q
iterations.

 After q iterations the intensity of the ith reconstructed projection ray is 
 $$ R_i^q = \sum_j p_{i,j} \rho_j^q , $$
and the density in each pixel is 
  $$\rho^{\sim q+1}_j= \rho_j^q + p_{i,j} {R_i - R_i^q \over N_i}
 ~~~~\mbox{with starting value}~~\rho^{\sim 0}_j=0 $$
where R$_i$ is the measured projection and,
\[ i= \left \{ \begin{array}{l}
              \mbox{m,if (q+1) is divisible m} \\
	      \mbox{the remainder of dividing (q+1)by m, otherwise}
		\end{array}
		\right. \] 
and,
    \[  \rho_j^q  = \left \{ \begin{array}{lll}
                               0,                   & \mbox{if $ \rho^{\sim q } \leq 0 $} \\
                                \rho^{\sim q}_j , & \mbox{if $0 \leq \rho^{ \sim q}_j \leq 1$} \\
			       1,                   & \mbox{if $ \rho^{\sim q}_j \geq 1$}
                               \end{array} 
                             \right. \]
It is necessary to determine when an iterative algorithm has converged
 to a solution which is optimal according to some criterion. We are 
using as the
criteria of convergence  the discrepancy between the measured and calculated
projection elements
$$D^q=\left\{ \frac{1}{m} \sum_{i=1}^m  \frac{\left( R_i - R_i^q \right)^2}{N_i} \right\}^{\frac{1}{2}} $$

We have added a third wire at 45 degree  
in  wire scanners in two places. Figure 5
compares the measured and reconstructed  profiles in the BLIP transfer line 
\cite{ahovi}
after the 1st octupole
and figure 6 shows the reconstructed 3D density distribution.
\begin{figure}
	\vspace*{2.2in}
	\includegraphics{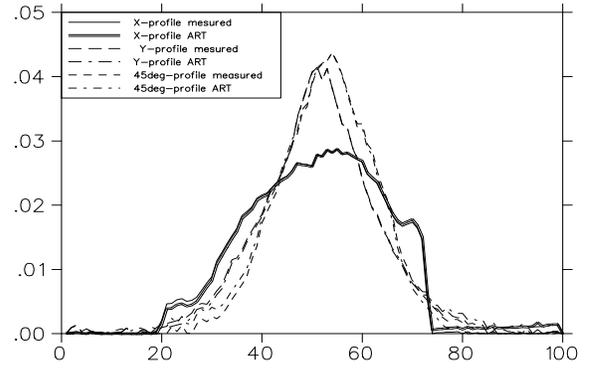}
	\caption{ Beam projection on x, y, and 45 degree planes.}
\end{figure}
\begin{figure}
	\vspace*{1.5in}
	\includegraphics{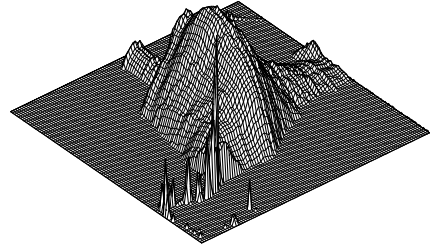}
	\caption{ Reconstructed 3D density distribution using ART.}
\end{figure}

\end{document}